\newcommand{\beq}{\begin{equation}}
\newcommand{\eeq}{\end{equation}}
\newcommand{\bea}{\begin{eqnarray}}

\newcommand{\eea}{\end{eqnarray}}

\newcommand{\ba}{\begin{array}}
\newcommand{\ea}{\end{array}}

\newcommand{\bkappa}{\mbox{\boldmath $\kappa$}}

\newcommand{\bb}{{\bf b}}
\newcommand{\br}{{\bf r}}

\newcommand{\bp}{{\bf p}}

\def\lsim{\mathrel{\rlap{\lower4pt\hbox{\hskip1pt$\sim$}}
    \raise1pt\hbox{$<$}}}         
\def\gsim{\mathrel{\rlap{\lower4pt\hbox{\hskip1pt$\sim$}}
    \raise1pt\hbox{$>$}}}         
\def\beq{\begin{equation}}
\def\eeq{\end{equation}}
\def\bea{\begin{eqnarray}}
\def\eea{\end{eqnarray}}

\documentclass[epj]{svjour}
\usepackage{graphics}
\begin{document}
\title{High Density QCD and Saturation of Nuclear Partons}
\author{I.P.Ivanov\inst{1,2},N.N. Nikolaev\inst{1,3},
W. Sch\"afer\inst{4}, B.G. Zakharov\inst{3} and V.R.
Zoller\inst{5} 
}                     
%
%
\institute{Institut f. Kernphysik, Forschungszentrum J\"ulich, D-52425 J\"ulich, Germany
\and 
Institute of Mathematics, Novosibirsk, Russia
\and 
L.D.Landau Institute for Theoretical Physics, Chernogolovka, Russia 
\and NORDITA, Blegdamsvej 17, DK-2100 Copenhagen \O, Denmark  \and 
Institute for Theoretical and Experimental Physics, Moscow, Russia}
\date{Received: date / Revised version: date}
%
\abstract{We review the recent finding of the two-plateau momentum 
distribution of sea quarks in deep inelastic scattering off
nuclei in the saturation regime. The diffractive plateau which
dominates for small $\bp$ measures precisely the momentum
distribution of quarks in the beam photon, the r\^ole of the
nucleus is simply to provide an opacity. The plateau for truly
inelastic DIS exhibits a substantial nuclear broadening of the 
momentum distribution. Despite this nuclear broadening, the 
observed final state and initial state sea quark densities do
coincide exactly. We emphasize how the saturated sea is
generated from the nuclear-diluted Weizs\"acker-Williams 
because of the anti-collinear splitting of gluons into sea quarks.
} 
\maketitle
\section{Introduction}
\label{intro}

The saturation of nuclear partons is one of hot issues
in high energy physics.
The interpretation of nuclear opacity in terms of a fusion
and saturation of nuclear partons has been introduced in
1975 \cite{NZfusion} way before the QCD parton model:
 the
Lorentz contraction of relativistic nuclei entails a spatial
overlap of partons with $x \lsim x_{A} \approx 1/R_A m_N$ from
different nucleons and the fusion of overlapping partons results
in the saturation of parton densities per unit area in the impact
parameter space. The pQCD link between nuclear opacity and
saturation has been considered  by Mueller \cite{Mueller1} and the
pQCD discussion of fusion of nuclear gluons has been revived by
McLerran et al. \cite{McLerran}. 

In this talk we review the recent consistent derivation of nuclear 
 parton densities in the opacity/saturation regime
\cite{NSSZsatur}. We demonstrate that despite the strong 
nuclear distortions the common wisdom  that in deep inelastic 
scattering  (DIS) the 
observed final state (FS) momentum distribution of struck partons
coincides with the initial state (IS) density of partons in the 
probed target holds for nuclei too. The key to this derivation 
is the Weizs\"acker-Williams (WW) glue of the relativistic 
nucleus as defined according to \cite{NSS}. We
pay a special attention to an important point that diffractive DIS
in which the target nucleus does not break and is retained in the
ground state, makes precisely 50 per cent of the total DIS events
\cite{NZZdiffr}. We point out that the saturated diffractive
plateau measures precisely the momentum distribution of
(anti-)quarks in the $q\bar{q}$ Fock state of the photon.  
We show how the anti-collinear splitting of WW
gluons into sea quarks gives rise to nuclear saturation of the sea
despite the substantial nuclear dilution of the WW glue.

\section{Nuclear distortions of the parton spectra}
\label{sec:1}

\begin{figure}
\resizebox{0.45\textwidth}{!}{%
\vspace*{13cm}
  \includegraphics{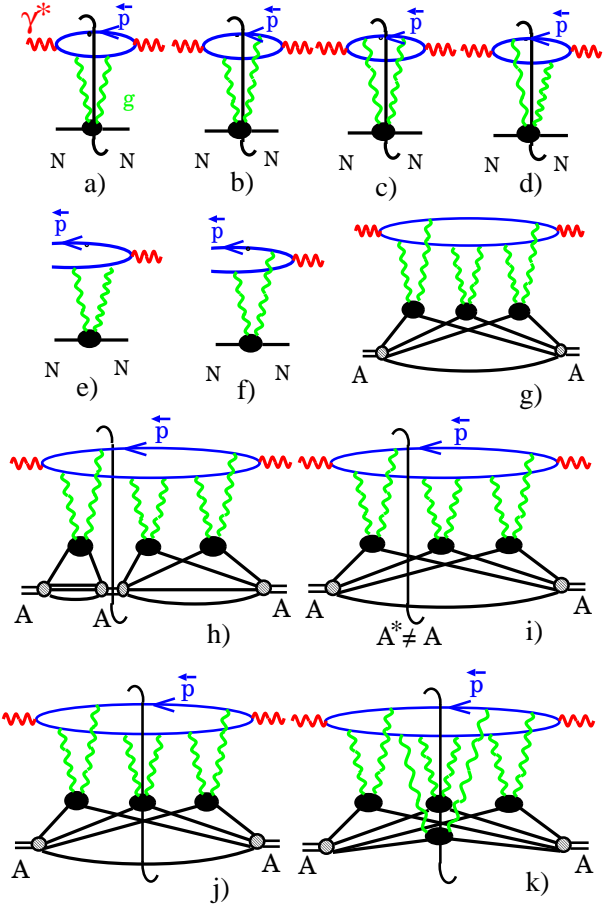}
}
\caption{The pQCD diagrams for inclusive (a-d) and
diffractive
(e,f) DIS off protons and nuclei
(g-k). Diagrams (a-d) show the unitarity cuts with color
excitation of the target nucleon, (g) - a generic multiple
scattering diagram for Compton scattering off nucleus, (h) - the
unitarity cut for a coherent diffractive DIS, (i) - the unitarity
cut for quasielastic diffractive DIS with excitation of the
nucleus $A^*$, (j,k) - the unitarity cuts for truly inelastic DIS
with single and multiple color excitation of nucleons of the
nucleus. }
\label{fig:2}       
\end{figure}

We illustrate our ideas on
an example of DIS at $x\sim x_A \ll 1$ which is dominated by
interactions of $q\bar{q}$ Fock states of the photon
\cite{NZZdiffr,NZ91,NZ92,NZ94,NZZlett} . The total cross
section for interaction of the color dipole $\br$ with the target
nucleon equals 
\bea \sigma(r)= \alpha_S(r) \sigma_0\int d^2\bkappa
f(\bkappa )\left[1 -\exp(i\bkappa \br )\right]\, , \label{eq:1}
\eea 
where the unintegrated glue of
the target nucleon is
\bea f(\bkappa ) = {4\pi \over
N_c\sigma_0}\cdot {1\over \kappa^4} \cdot {\partial G \over
\partial\log\kappa^2}\, , ~~~~\int d^2\bkappa  f(\bkappa )=1\, .
\label{eq:2} 
\eea
For DIS off a free nucleon target, see figs. 1a-1d, the 
spectrum of the FS
quark  prior the hadronization,
\bea
{d\sigma_N \over d^2\bp dz} =
{\sigma_0\alpha_S(\bp^2) \over 2(2\pi)^2}
 \int d^2\bkappa f(\bkappa )\nonumber\\
\left|\langle \gamma^*|\bp\rangle - \langle \gamma^*|\bp-\bkappa \rangle\right|^2\, ,
\label{eq:3}
\eea
where $\bp$ is the transverse momentum, and $z$ the Feynman variable,
coincides, upon the $z$ integration, with
the conventional IS unintegrated $\bp$
distribution of partons in the target.
Notice that the target nucleon is color-excited and
there is no rapidity gap in the FS.

In  DIS off nuclei one must distinguish the three principal
processes with distinct unitarity cuts of the forward Compton
amplitude (fig.~1g): the coherent diffraction dissociation {\sl
(D)} of the photon (fig.~1h), quasielastic diffraction
dissociation {\sl (qel)} followed by excitation and breakup of the
target nucleus  (fig.~1i) - in both of them there is no color flow
between the photon debris and the nucleus - , and the truly
inelastic {\sl (in)} DIS with color excitation of nucleons of the
target nucleus (fig.~1j,k). 

The technical trick with separation 
of the color dipole-nucleon $S$-matrix into the color-octet 
and color-singlet components, and the derivation of the full
two-parton momentum spectrum are found in \cite{NSSZsatur},
here we only cite the single-parton spectrum
\bea 
{d \sigma_{in}\over d^2\bp dz }   =  {1
\over (2\pi)^2}\int d^2\bb
 \int d^2\br' d^2\br
\exp[i\bp(\br'-\br)]\nonumber\\
\times \Psi^*(\br')\Psi(\br) 
\times \left\{\exp[-{1\over 2}\sigma(\br-\br')T(\bb)]\right.\\
\left. -
\exp[-{1\over 2}[\sigma(\br)+\sigma(\br')]T(\bb)]\right\}\, .\nonumber
\label{eq:10} \eea 
Evidently, the dependence of nuclear
attenuation factors on $\br,\br'$ shall distort strongly the
observed momentum distribution of quarks.

\section{WW glue of nuclei}

In the further parton model interpretation of this spectrum 
 we resort to the NSS
representation \cite{NSS}
\bea
\Gamma_A(\bb,\br)= 1-\exp\left[-{1\over 2}\sigma(r)T(\bb)\right]\nonumber\\
=
 \int d^2\bkappa  \phi_{WW}(\bkappa )[1-\exp(i
\bkappa \br) ] \, ,
\label{eq:11}
\eea
where, driven by a  close analogy to (\ref{eq:1}),(\ref{eq:2})
in terms of $f(\bkappa)$,  we interpret
\beq
 \phi_{WW}(\bkappa ) =
\sum_{j=1}^{\infty} \nu_A^j(\bb) \cdot {1 \over j!}
f^{(j)}(\bkappa ) \exp\left[-\nu_A(\bb)\right] \label{eq:12} 
\eeq
as the WW unintegrated glue of a nucleus
per unit area in the impact parameter plane. Here $$\nu_A(\bb)= {1 \over
2}\alpha_S(r)\sigma_0T(\bb)$$ defines the nuclear opacity and the
$j$-fold convolutions \beq f^{(j)}(\bkappa )= \int \prod_{i=1}^j
d^2\bkappa _{i} f(\bkappa _{i}) \delta(\bkappa -\sum_{i=1}^j
\bkappa _i) \label{eq:14} \eeq describe the contribution to the
diffractive amplitudes from $j$ split pomerons \cite{NSS}.

A discussion of the nuclear antishadowing property of the hard WW
glue is found in \cite{NSS}. A somewhat involved analysis of the
properties of the convolutions (\ref{eq:14}) in the soft region
shows that they develop a plateau-like behaviour with the width of
the plateau which expands $\propto j$. The gross features of the
WW nuclear glue in the soft region are well reproduced by \bea
\phi_{WW}(\bkappa) \approx  {1\over \pi}  {Q_A^2 \over (\bkappa^2
+Q_{A}^2)^2}\, , \label{eq:15} \eea where the saturation scale $
Q_A^2 =  \nu_A(\bb)  Q_0^2 \propto A^{1/3}\, $. 
Notice the nuclear dilution of soft WW glue, $\phi_{WW}(\bkappa) \propto 1/Q_A^2 \propto
A^{-1/3}$.
 The soft
parameters $Q_0^2$ and $\sigma_0 $ are related to the integrated
glue of the proton in the soft region:
$$
Q_{0}^2\sigma_0 \sim {2\pi^2 \over N_c} G_{soft}\,, ~~~G_{soft}\sim 1\,.
$$

\section{An exact equality of the IS and FS parton densities}

Making use of the NSS representation, the total
nuclear photoabsorption cross section can be cast in the form 
\bea
\sigma_{A}  = \int d^2\bb\int dz 
\int {d^2\bp\over (2\pi)^2}\nonumber\\ 
\int
d^2\bkappa\phi_{WW}(\bkappa) \left|\langle \gamma^* |\bp\rangle -
\langle \gamma^* |\bp-\bkappa\rangle \right|^2 \label{eq:16} 
\eea
which has a profound  semblance to (\ref{eq:3}) and one is tempted
to take the differential form of (\ref{eq:16}) as a definition of
the IS sea quark density in a nucleus: \bea {d\bar{q}_{IS} \over
d^2\bb d^2\bp} = {1\over 2}\cdot{Q^2 \over 4\pi^2 \alpha_{em}}
\cdot{d\sigma_A \over d^2\bb d^2\bp}\, . \label{eq:17} \eea 
Remarkably, in
terms of the NSS-defined
WW nuclear glue, all intranuclear multiple-scattering
diagrams of fig.~1g sum up to precisely the same four diagrams
fig.~1a-1d as in DIS off free nucleons.  Although $\bp$ emerges
here just as a formal Fourier parameter, we shall demonstrate that
it can be identified with the momentum of the observed final state
antiquark.

Indeed, making use of the NSS representation, after
some algebra one finds 
 \bea {d \sigma_{in}\over d^2\bb d^2\bp dz } =
{1 \over (2\pi)^2}\nonumber\\
\times \left\{ \int  d^2\bkappa
\phi_{WW}(\bkappa)\left|\langle \gamma^* |\bp\rangle -
\langle \gamma^* |\bp-\bkappa\rangle \right|^2 \right. \nonumber\\
 - \left. \left|\int d^2\bkappa\phi_{WW}(\bkappa) (\langle
\gamma^* |\bp\rangle - \langle \gamma^*
|\bp-\bkappa\rangle )\right|^2\right\} \label{eq:19} \\
{d \sigma_{D}\over d^2\bb d^2\bp dz }   = {1 \over (2\pi)^2}\nonumber\\
\times\left|\int d^2\bkappa\phi_{WW}(\bkappa) (\langle \gamma^*
|\bp\rangle - \langle \gamma^* |\bp-\bkappa\rangle)\right|^2 \, .
\label{eq:20} \eea 
As far as diffraction is concerned, the analogy
between (\ref{eq:20}) and its counterpart for free nucleons
\cite{NSS,NZ92,NZsplit}, and nuclear WW glue $\phi_{WW}(\bkappa)$
and $f(\bkappa)$ thereof, is complete. Putting the inelastic and
diffractive components of the FS quark spectrum together, we
evidently find the FS parton density which exactly coincides with
the IS parton density (\ref{eq:17}) such that $\bp$ is indeed the
transverse momentum of the FS sea quark. The interpretation of
this finding is not trivial, though.

\section{The two-plateau spectrum of sea quarks}

Consider first the domain of $\bp^2 \lsim Q^2 \lsim Q_A^2$ such
that the nucleus is opaque for all color dipoles in the photon.
Hereafter we assume that the saturation scale $Q_A^2$ is so large
that $\bp^2,Q^2$ are in the pQCD domain and neglect the quark
masses. In this regime the nuclear counterparts of the crossing
diagrams of figs. 1b,d,f  can be neglected. Then, in the
classification of \cite{NSS}, diffraction will be dominated by the
contribution from the Landau-Pomeranchuk diagram of fig.~1e with
the result 
\bea \left. {d\bar{q}_{FS} \over d^2\bb
d^2\bp}\right|_D
\approx {1\over 2}\cdot{Q^2 \over 4\pi^2 \alpha_{em}}
\int
dz  \nonumber\\
\times
\left| \int d^2\bkappa\phi_{WW}(\bkappa) \right|^2
\left|\langle \gamma^* |\bp\rangle\right|^2  \approx {N_c \over
4\pi^4}\, . \label{eq:21} \eea 
Remarkably, diffractive DIS measures the momentum distribution of
quarks and antiquarks in the $q\bar{q}$ Fock state of the photon.
This result, typical of the Landau-Pomeranchuk
mechanism, is a completely generic one and would hold for any beam
particle such that its coupling to colored partons is weak. In
contrast to diffraction off free nucleons
\cite{NZ92,NZsplit,GNZcharm}, diffraction off opaque nuclei is
dominated by the anti-collinear splitting of hard gluons into soft
sea quarks, $\bkappa^2 \gg \bp^2$. Precisely for this reason one
finds the saturated FS quark density, because the nuclear dilution
of the WW glue is compensated for by the expanding plateau. The
result (\ref{eq:21}) has no counterpart in DIS off free nucleons
because diffractive DIS off free nucleons is negligibly small even
at HERA, $\eta_D \lsim $ 6-10 \%.

The related analysis of the FS quark density for truly inelastic
DIS in the  same domain of $\bp^2 \lsim Q^2 \lsim Q_A^2$ gives
\bea \left.{d\bar{q}_{FS} \over d^2\bb d^2\bp}\right|_{in} =
{1\over 2}\cdot{Q^2 \over 4\pi^2 \alpha_{em}} \cdot\int dz \nonumber\\
\times\int
d^2\bkappa \phi_{WW}(\bkappa)
\left|\langle \gamma^* |\bp-\bkappa\rangle \right|^2 \nonumber\\
 =
{Q^2 \over 8\pi^2 \alpha_{em}}\phi_{WW}(0)
\int^{Q^2} d^2\bkappa \int dz \left|\langle \gamma^* |\bkappa\rangle \right|^2
 \nonumber\\\approx 
{N_c \over 4\pi^4}\cdot {Q^2 \over Q_A^2}\cdot\theta(Q_A^2-\bp^2)\, .
\label{eq:22}
\eea
It describes final states with color excitation of a nucleus,
but as a function of the photon wave
function and nuclear WW gluon distribution it is completely
different from (\ref{eq:3}) for free nucleons. The $\theta$-function simply
indicates that the plateau for inelastic DIS extends up to $\bp^2
\lsim Q^2_A$.
For $Q^2 \ll Q_A^2$ the inelastic plateau
contributes little to the transverse momentum distribution of
soft quarks, $\bp^2 \lsim Q^2$, but the inelastic plateau extends way beyond
$Q^2$ and its integral contribution to the spectrum of FS
quarks is exactly equal to that from diffractive DIS. Such a  two-plateau
structure of the FS quark spectrum is a new finding and has not been
considered before.

Now notice, that  in the opacity regime the diffractive FS parton
density coincides with the contribution $\propto |\langle
\gamma^*|\bp\rangle|^2$ to the IS sea parton density from the
spectator diagram 1a, whereas the FS parton density for truly
inelastic DIS coincides with the contribution to IS sea partons
from the diagram of fig.~1c. The contribution from the crossing
diagrams 1b,d is negigibly small.

Our results (\ref{eq:21}) and (\ref{eq:22}), especially nuclear broadening
and unusually strong $Q^2$ dependence of the FS/IS parton density from
truly inelastic DIS, demonstrate clearly a distinction between diffractive
and inelastic DIS. Our considerations can readily be extended to the
spectrum of soft quarks, $\bp^2 \lsim Q_A^2$, in hard photons, $Q^2 \gsim Q_{A}^2$.
In this case the result (\ref{eq:21}) for diffractive DIS is retained,
whereas in the numerator of the result (\ref{eq:22}) for truly inelastic
DIS one must substitute $Q^2 \to Q_{A}^2$, so that in this case
$dq_{FS}|_{D} \approx dq_{FS}|_{in}$ and $dq_{IS} \approx
2dq_{FS}|_{D}$. The evolution
of soft nuclear sea, $\bp^2 \lsim Q_{A}^2$, is entirely driven by
an anti-collinear splitting of the NSS-defined WW nuclear glue into the sea
partons.

The early discussion of the FS quark density in the saturation
regime is due to Mueller \cite{Mueller}. Mueller focused on $Q^2
\gg Q_A^2$ and discussed  neither a distinction between
diffractive and truly inelastic DIS nor a $Q^2$ dependence and
broadening (\ref{eq:15}) for truly inelastic DIS at $Q^2 \lsim
Q_A^2$.

\section{More signals of saturation in diffractive DIS}

The flat $\bp^2$ distribution of forward $q,\bar{q}$ jets 
in truly inelastic DIS in the saturation regime must be contrasted 
to the $\propto G(\bp^2)/\bp^2$ spectrum for the free nucleon
target. In the diffractive DIS the saturation gain is much more
dramatic: flat $\bp^2$ distribution of forward $q,\bar{q}$ jets 
in diffractive DIS in the saturation regime must be contrasted 
to the $\propto 1/(\bp^2)^2$ spectrum for the free nucleon
target \cite{NZ92,NZsplit}. 
In the exclusive diffractive DIS, i.e., the vector meson
production, the relevant hard scale for the proton target 
equals $\bar{Q}^2 \approx {1\over 4}(Q^2+m_V^2)$
and the transverse cross section has been predicted to behave
as \cite{NNZscan}
\beq
\sigma_T \propto G^2(x,\bar{Q}^2)(\bar{Q}^2)^{-4}
\eeq
At $\bar{Q}^2 > Q_A^2$ the same would hold for nuclei too, 
but in the opposite case of $\bar{Q}^2 < Q_A^2$ the 
$\bar{Q}^2$-dependence is predicted to change to 
\beq
\sigma_T \propto G^2(x,\bar{Q}^2)(Q_A^2)^{-2}
(\bar{Q}^2)^{-2}
\eeq

\section{Conclusions}

 We reported a derivation of the FS parton
spectrum. Our result (\ref{eq:10}) summarizes in an elegant way
intranuclear distortions due to multiple diffractive rescatterings
and color excitations of the target nucleus. In conjunction with
the NSS definition of the WW glue of the nucleus, eqs.~
(\ref{eq:11}), (\ref{eq:12}), it gives an explicit form of the FS
parton densities. The two-plateau FS quark density with the strong
$Q^2$ dependence of the plateau for truly inelastic DIS has not
been discussed before. A comparison with the IS nuclear parton
densities which evolve from the NSS-defined WW nuclear glue shows
an exact equality of the FS and IS parton densities. The
plateau-like saturated nuclear quark density is suggestive of the
Fermi statistics, but our principal point that for any projectile
which interacts weakly with colored partons the saturated density
measures the momentum distribution in the $q\bar{q} , gg,...$ Fock
state of the projectile disproves the Fermi-statistics
interpretation. The spin and color multiplet of colored partons
the photon couples to is completely irrelevant, what only counts
is an opacity of heavy nuclei. The anti-collinear splitting of WW
nuclear glue into soft sea partons is a noteworthy feature of the
both diffractive DIS and IS sea parton distributions. The
emergence of a saturated density of IS sea partons from the
nuclear-diluted WW glue is due to the nuclear broadening of the
plateau (\ref{eq:15}). Because the predominance of diffraction is
a very special feature of DIS \cite{NZZdiffr}, one must be careful
with applying the IS parton densities to, for instance, nuclear
collisions, in which diffraction wouldn't be of any significance.

One can go one step further and consider interactions with the
opaque nucleus of the $q\bar{q}g$ Fock states of the photon.
Then the above analysis can be extended to $x \ll x_A$  and
the issue of the $x$-dependence of the saturation scale
$Q_A^2$ can be addressed following the discussion in \cite{NZ94}.
We only mention here that as far as diffraction and IS parton
densities are concerned, the NSS-defined WW glue
remains a useful concept and the close correspondence between
$\phi_{WW}(\bkappa)$ for the nucleus and $f(\bkappa)$ for the nucleon is
retained.

This work has been partly supported by DAAD and
Nordita, the INTAS grants 97-30494
\& 00-00366 and DFG grant 436RUS17/119/01.

%
%

\end{document}